\documentclass[12pt,preprint]{aastex}

\begin{document}

\title{\bf DIFFERENCE BETWEEN SPATIAL DISTRIBUTIONS OF THE H$\alpha$ KERNELS 
AND HARD X-RAY SOURCES IN A SOLAR FLARE}

\author{Ayumi Asai \altaffilmark{1}, 
Satoshi Masuda \altaffilmark{2}, 
Takaaki Yokoyama \altaffilmark{3}, 
Masumi Shimojo \altaffilmark{3}, 
Hiroaki Isobe \altaffilmark{1}, 
Hiroki Kurokawa \altaffilmark{1}, and 
Kazunari Shibata \altaffilmark{1}}
\email{asai@kwasan.kyoto-u.ac.jp}
\altaffiltext{1}{
Kwasan and Hida Observatories, Kyoto University, Yamashina-ku, 
Kyoto 607-8471, JAPAN}
\altaffiltext{2}{
Solar-Terrestrial Environment Laboratory, Nagoya University, Toyokawa, 
Aichi 442-8507, JAPAN}
\altaffiltext{3}{
Nobeyama Radio Observatory, Minamisaku, Nagano, 384-1305, JAPAN}

\begin{abstract}

We present the relation of the spatial distribution of H$\alpha$ kernels 
with the distribution of hard X-ray (HXR) sources seen during the 2001 
April 10 solar flare.
This flare was observed in H$\alpha$ with the {\it Sartorius} telescope at 
Kwasan Observatory, Kyoto University, and in hard X-rays (HXRs) with the 
Hard X-ray Telescope (HXT) onboard {\it Yohkoh}.
We compared the spatial distribution of the HXR sources with that of the 
H$\alpha$ kernels.
While many H$\alpha$ kernels are found to brighten successively during 
the evolution of the flare ribbons, only a few radiation sources are seen in 
the HXR images.
We measured the photospheric magnetic field strengths at each radiation source 
in the H$\alpha$ images, and found that the H$\alpha$ kernels accompanied by 
HXR radiation have magnetic strengths about 3 times larger than those 
without HXR radiation.
We also estimated the energy release rates based on the magnetic reconnection 
model.
The release rates at the H$\alpha$ kernels with accompanying HXR sources are 
16-27 times larger than those without HXR sources.
These values are sufficiently larger than the dynamic range of HXT, 
which is about 10, so that the difference between the spatial distributions 
of the H$\alpha$ kernels and the HXR sources can be explained.

\end{abstract}


\section{INTRODUCTION}

In the impulsive phase of a solar flare, precipitations of nonthermal 
electrons from the corona generate radiation from denser layers, such as 
the transition region and/or the upper chromosphere.
This radiation is often observed in hard X-rays (HXRs) or microwaves.
Precipitations of nonthermal electrons also cause radiation sources 
in H$\alpha$ because of rapid thermalization or other mechanisms.
Therefore, H$\alpha$ kernels and HXR sources show a high correlation in their 
locations and light curves \citep{Kita90}.
However, the difference between the spatial distributions of H$\alpha$ kernels 
and HXR sources is also well known.
H$\alpha$ images sometimes show elongated brightenings, called H$\alpha$ flare 
ribbons, with many H$\alpha$ kernels within them. 
The size of elemental H$\alpha$ kernels is considered to be about 
$1^{\prime\prime}$ or even smaller \citep{Kuro86}, which is larger than the 
spatial resolution achieved with H$\alpha$ instruments 
($\sim 0^{\prime\prime}_{\,\cdot}2$) but is smaller than that with HXR 
instruments of about $5^{\prime\prime}$.
On the other hand, HXR images show very few sources, sometimes only one.
HXR sources are accompanied by H$\alpha$ kernels in many cases, 
but many H$\alpha$ kernels are not accompanied by HXR sources.
The only exception to this ``lack of radiation sources in HXRs'' that is 
known, is the Bastille Day event on 2000 July 14 \citep{Masu01}.
This event shows a clear two-ribbon structure in HXRs such as in H$\alpha$.

This difference of spatial distributions may be explained by the difference 
in radiation mechanisms between HXRs and H$\alpha$.
The HXR intensity is proportional to the number of accelerated electrons, 
and is thought to be proportional to the energy release rate 
\citep{Hud91,Wu86}.
Therefore, only compact regions where the largest energy release occurred are 
observable as HXR sources.
On the other hand, the mechanisms for H$\alpha$ radiation are much more 
complicated than those for HXR radiation, and to derive the effect of 
electrons is quite difficult \citep{Ric83,Can84}.
Some weaker H$\alpha$ kernels may be caused by a secondary effect of 
precipitation or thermal conduction.
However, as we mentioned above, the light curve of each H$\alpha$ kernel has 
a high correlation with that of the total HXR intensity, even if 
their intensity is not so strong and they do not have an HXR counterpart.
We suggest that the difference between the spatial distributions of H$\alpha$ 
kernels and HXR sources is caused by the low dynamic range of the HXR data.
In the HXR images only the strongest sources are seen, and the weaker sources 
are buried in the noise.
We use the HXR data taken with the hard X-ray telescope (HXT; Kosugi et al. 
1991) onboard {\it Yohkoh} (Ogawara et al. 1991) in this paper.
The dynamic range of the HXT images is about 10.
Therefore, if the released energy at the H$\alpha$ kernels associated 
{\it with} HXR sources is at least 10 times larger than that at the H$\alpha$ 
kernels {\it without} HXR sources, then the difference of appearance can be 
explained. 

In this paper, we measured the photospheric magnetic field strengths at 
each radiation source seen in H$\alpha$ images which have much higher 
spatial resolution than HXR images.
We also estimated the energy release rates based on the magnetic reconnection 
model \citep{Iso02} at each radiation source.
Then we compared the energy release rates with the spatial distribution of 
radiation sources in an HXR image since they suggest the site where the strong 
energy release occurred.
To examine the difference of the amount of the released magnetic energy, 
we measured the photospheric magnetic field strengths at each radiation source.
In \S 2 we summarize the observational data and the results.
In \S 3 we discuss the amount of energy release at each radiation source.
In \S 4 the summary and the conclusion are given.

\section{OBSERVATIONS AND RESULTS}

We observed a large two-ribbon flare (X2.3 on the GOES scale) which 
occurred in the NOAA Active Region 9415 (S22$^{\circ}$, W01$^{\circ}$) at 
05:10 UT, 2001 April 10 with the Sartorius Refractor Telescope 
({\it Sartorius}) at Kwasan Observatory, Kyoto University.
The highest temporal and spatial resolutions of the {\it Sartorius} data are 
1 s and $1^{\prime\prime}_{\,\cdot}2$, respectively.
The details of the flare were reported in several papers 
\citep{Asa02,Pike02,Fol01}.
To compare the locations of the HXR radiation sources with those of the 
H$\alpha$ kernels, we used the HXR data taken with HXT, whose temporal and 
spatial resolutions are 0.5 s and $5^{\prime\prime}$, respectively.
We also used a magnetogram obtained at 05:18 UT with the Michelson Doppler 
Imager (MDI; Scherrer et al. 1995) onboard the 
{\it Solar and Heliospheric Observatory} ({\it SOHO}; Domingo, Fleck, and 
Poland 1995) to measure photospheric magnetic field strengths.
The spatial resolution of the MDI image is about 
$3^{\prime\prime}_{\,\cdot}9$.

\begin{figure}
\plotone{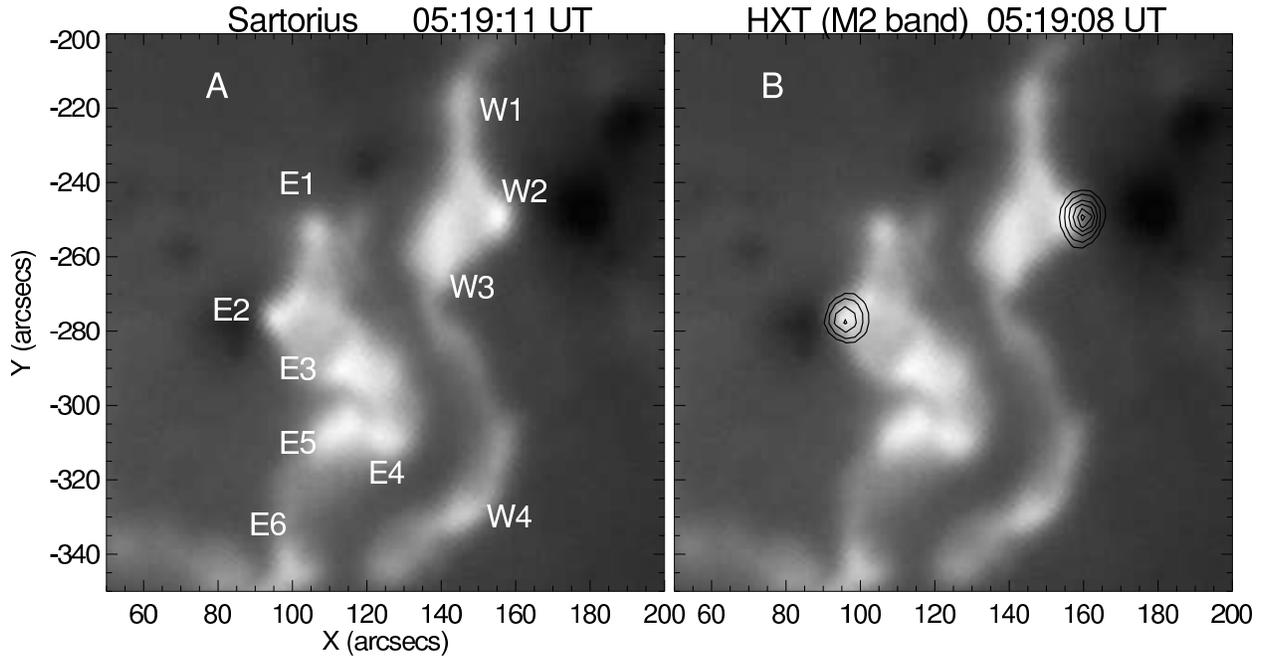}
\caption{
H$\alpha$ image of the flare at 05:19 UT.
Solar north is up, and west is to the right.
Panel A is an H$\alpha$ image taken with ({\it Sartorius}) in which 
ten H$\alpha$ bright kernels are numbered from E1 to E6 and from W1 to W4.
Panel B is the H$\alpha$ image overlaid with the HXR contour image to compare 
the spatial distribution of H$\alpha$ kernels with that of the HXR sources.
Contour levels are 95\%, 80\%, 60\%, 40\%, 20\% and 10\% of the peak intensity.
\label{fig1}}
\end{figure}

Figure \ref{fig1} shows the images of the flare at 05:19 UT in H$\alpha$ 
and HXR.
We overlaid the HXR contour image on the H$\alpha$ image (in panel B of 
Fig. \ref{fig1}), to compare the spatial distribution of radiation sources 
in these two wavelengths.
There are many H$\alpha$ kernels inside the flare ribbons.
In panel A of Figure 1 we can identify ten H$\alpha$ kernels by eyes, 
which are numbered from E1 to E6 and from W1 to W4.
Here, ``E'' and ``W'', mean the eastern and the western flare ribbons, 
respectively.

On the other hand, the HXR image (panel B in Fig. \ref{fig1}) shows 
only two sources even at the contour level of 10\% of the peak intensity.
These HXR sources are associated with the H$\alpha$ kernels E2 and W2.
We confirmed that those H$\alpha$ kernels, E2 and W2, are conjugate 
footpoints in a previous paper \citep{Asa02}.
The soft X-ray images of the flares taken with the Soft X-ray Telescope 
onboard {\it Yohkoh} also show the flare loops which connect E2 and W2.
Such differences between the spatial distributions of the H$\alpha$ kernels 
and the HXR sources may be caused by the differences in energy release rates 
at each radiation source.
We measured the magnetic field strengths at each H$\alpha$ kernel, 
both with and without HXR sources, and estimated their energy release rates.
Then we compared these rates with the spatial distribution of the HXR sources.
The estimation is discussed in the next section.
Here, we examine the relation between the photospheric magnetic field 
strength and the radiation sources.

\begin{figure}
\plotone{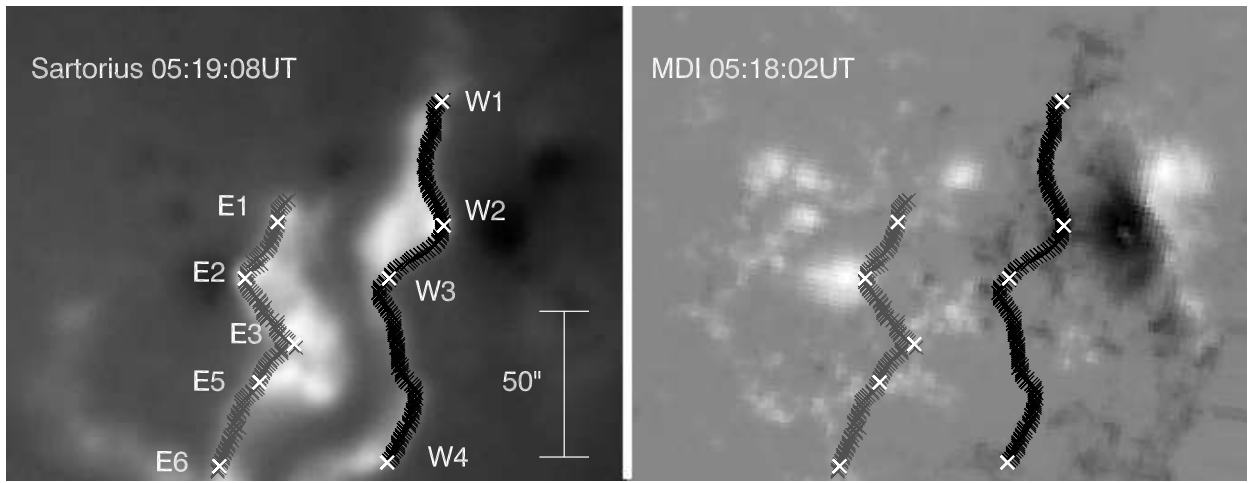}
\caption{
H$\alpha$ images ({\it Sartorius}) and photospheric magnetogram (MDI).
Celestial north is up, and west is to the right.
The outer edges of both the flare ribbons are plotted by cross signs.
{\it Gray} and {\it black} plots show the east (positive) and west (negative) 
magnetic polarity, respectively.
\label{fig2}}
\end{figure}

Figure \ref{fig2} shows the H$\alpha$ image in which the flare ribbons are 
clearly seen ({\it left} panel), 
and the photospheric magnetogram of the same region obtained with MDI 
({\it right} panel).
The outer edges of the H$\alpha$ flare ribbons are plotted by cross signs 
in both the panels of Figure \ref{fig2}.
The magnetic reconnection model indicates that the newest energy release 
occurs on the outermost flare loops, and the H$\alpha$ kernels and HXR sources 
appear at the footpoints of the flare loops, say, at the fronts of the 
H$\alpha$ flare ribbons.
Therefore, we measured the photospheric magnetic field strengths along the 
outer edges of the flare ribbons.
The source E4 is excluded from the discussion, because it is not located 
on the ribbon-front.

\begin{figure}
\plotone{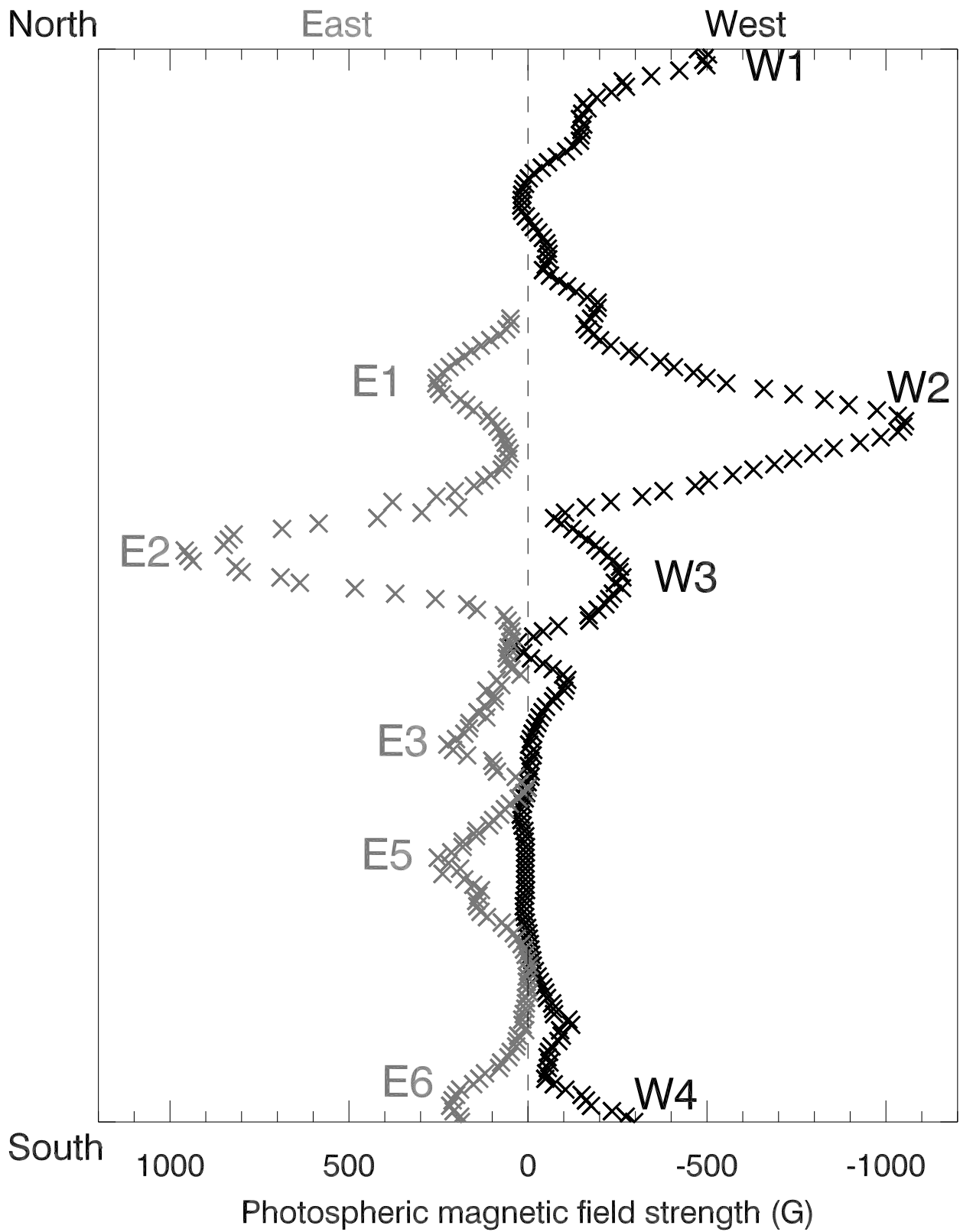}
\caption{
Magnetic field strength along the outer edges of both the flare ribbons 
(see Fig. \ref{fig2}).
{\it Left gray} plot shows the east (positive) ribbon, 
{\it right black} plot shows the west (negative) one.
Note E1 - E6 (except for E4) and W1 - W4 indicate the position of H$\alpha$ 
kernels.
E2 and W2 are associated with HXR sources.
\label{fig3}}
\end{figure}

Figure \ref{fig3} shows the photospheric magnetic field strengths along the 
outer edges of both the H$\alpha$ flare ribbons.
The {\it left gray} plot shows the magnetic strengths for the east ribbon with 
positive magnetic polarity, and the {\it right black} plot is for 
the west ribbon with negative magnetic polarity.
The magnetic field strengths at the H$\alpha$ kernels are higher than 
in the other non-kernel regions, 300 G on average, 
and those at the HXR sources (E2 and W2) are especially high ($\sim$ 1000 G).
We summarize the magnetic field strengths at each H$\alpha$ kernel in Table 1.
The photospheric magnetic field strengths of the H$\alpha$ kernels accompanied 
by HXR sources are about 3 times larger than the other H$\alpha$ kernels.

\section{ENERGY RELEASE RATE}

In this section, we examine the energy release rates in the flare ribbons, 
and discuss the relation between the rates and the spatial distribution of 
the H$\alpha$ kernels and HXR sources.
We assume that the HXR intensity observed with the HXT is proportional to the 
energy release rate ${dE}/{dt}$ due to magnetic reconnection.
The energy release rate is written as the product of the Poynting flux into 
the reconnection region ${(4 \pi)^{-1}}{B_{\rm corona}^{2}}{v_{\rm in}}$ and 
the area of the reconnection region $A$ \citep{Iso02} as follows; 
\begin{equation}
\frac{dE}{dt} = \frac{B_{\rm corona}^{2}}{4 \pi}v_{\rm in}A ,
\end{equation}
where $B_{\rm corona}$ is the magnetic field strength in the corona and 
$v_{\rm in}$ is the inflow velocity into the reconnection region.
For simplicity, we assume that the area of the reconnection region does not 
change much during the flare and is independent of the magnetic field 
strength.
Therefore, the energy release rate is simply written as 
${dE}/{dt} \propto B_{\rm corona}^{2}v_{\rm in}$. 
Here, we use magnetic field strengths at the the photosphere $B_{\rm photo}$ 
instead of those at the corona $B_{\rm corona}$ to estimate the energy release 
rates, since to measure $B_{\rm corona}$ directly is difficult.
We assume that $B_{\rm corona}$ is proportional to $B_{\rm photo}$ in 
the same ratio all over the flaring region.
If this assumption is true, the differences of the energy release rates 
estimated with $B_{\rm photo}$ are the same as those with $B_{\rm corona}$.
From now on, we estimate the energy release rate by using the photospheric 
magnetic field strength, and we write it simply as $B$.

If $v_{\rm in}$ has no dependence on $B$, the energy release rate is directly 
proportional to the square of the magnetic field strength, 
${dE}/{dt} \propto B^{2}$.
Since the H$\alpha$ kernels accompanied by HXR radiation have magnetic field 
strengths about 3 times larger than those without the HXR radiation, 
the energy release rates at the former are about 9 times larger than those 
at the latter.
The difference in the energy release rates is just comparable to the dynamic 
range of the HXT ($\sim$ 10), but is not enough for weaker radiation 
sources to be buried in noise.
At other H$\alpha$ kernels, HXR sources should be observed.
Therefore, under this assumption the spatial distribution of radiation 
sources cannot be explained.

This result suggests that $v_{\rm in}$ may have some dependence on $B$ as 
some authors have suggested.
\citet{Swe58} and \citet{Par57} give the relation 
$v_{\rm in} = {R_m}^{-1/2}v_{\rm A}$, where $R_m$ is the magnetic Reynolds 
number, and $v_{\rm A}$ is the Alfv\'{e}n velocity.
Here, $v_{\rm A}$ is expressed as 
$v_{\rm A} = (4 \pi \rho)^{-1/2}B_{\rm corona} \propto B_{\rm corona}$, 
where $\rho$ is the mass density.
$R_m$ is defined as $R_m = L v_{\rm A}/\eta \propto B$, where $L$ and 
$\eta$ are the characteristic length of the flare region and the magnetic 
diffusivity, respectively.
Therefore, we can derive the relation $v_{\rm in} \propto B^{1/2}$, and then 
${dE}/{dt} \propto B^{5/2}$.
On the other hand, \citet{Pet64} suggests that $v_{\rm in}$ hardly depends 
on $R_m$, and indicates 
$v_{\rm in} \propto ({\rm log}R_m)^{-1}v_{\rm A} \approx v_{\rm A} \propto B$.
Hence, we can derive the relation ${dE}/{dt} \propto B^{3}$.
Since $B$ at the HXR sources is 3 times larger, 
these two models of magnetic recconection predict that the energy release is 
16 and 27 times stronger for the H$\alpha$ sources accompanied by HXR sources 
than for the other H$\alpha$ kernels, respectively.
This is sufficiently larger than the dynamic range of HXT, and 
the difference between the spatial distributions can be explained.

\section{SUMMARY AND CONCLUSION}

We have examined the difference between the spatial distributions of 
H$\alpha$ kernels and HXR sources.
The HXR sources indicate where large energy release has occurred, 
while the H$\alpha$ kernels show the precipitation sites of 
nonthermal electrons with higher spatial resolution.
We measured the photospheric magnetic field strength at each H$\alpha$ 
kernel, and found that the magnetic field strengths at the H$\alpha$ kernels 
accompanied by HXR sources are about 3 times higher than those at the 
other H$\alpha$ kernels (without any HXR sources).
We also estimated their energy release rates, by using the photospheric 
magnetic field strengths and by considering the dependence of $v_{i}$ on $B$ 
as derived by some authors (e.g. Sweet 1958; Parker 1957; Petschek 1964).
The estimated energy release rates at the HXR sources are large enough to 
explain the difference of appearance of the H$\alpha$ and HXR images.

The gap in our understanding of large scale structure such as H$\alpha$ flare 
ribbons and compact radiation sources has been well known not only in HXRs, 
but also in microwaves.
Even in SXRs, flare loops often appear only in a part of the whole flaring 
region and/or seem to connect only a few radiation sources.
The lack of radiation sources does not mean that no energy release occurs 
there.
We just cannot ``observe'' the radiation because of the dynamic ranges 
of the instrument for those wavelengths.

We compared the photospheric magnetic field strenghts at each H$\alpha$ 
kernel with the spatial distribution of HXR sources, and found that HXR 
sources appear at strong magnetic regions.
However, our result was based on only one event, 
and the statistical studies about the relation between HXR intensities and 
magnetic field strengths are needed.
We will be able to perform more detailed analysis in the near future 
by using HXR images, with a higher dynamic range, obtained with 
the {\it Reuven Ramaty High Energy Solar Spectroscopic Imager} ({\it RHESSI}).

\acknowledgments

We firstly acknowledge an anonymous referee for his/her useful comments and 
suggestions.
We wish to acknowledge all the members of Kwasan observatory for their support
during our observation, especially Ms. M. Kamobe.
We also thank Dr. D. H. Brooks for his careful reading and correction of our 
paeper.
We wish to thank Drs. S. Tanuma, N. V. Nitta, J. Sato, T. Kosugi, 
K. Shibasaki, for fruitful discussions and their helpful comments.
We again would like to thank them for their encouragement.
The {\it Yohkoh} satellite is a Japanese national project, launched and 
operated with by ISAS, and involving many domestic institutions, with 
multilateral international collaboration with the US and the UK.
{\it SOHO} is a mission of international cooperation between the European 
Space Agency (ESA) and NASA.

\clearpage

\begin{table}
\begin{center}
Table 1: Photospheric magnetic field strengths at H$\alpha$ kernels\\

\begin{tabular}{lcc}\tableline\tableline
H$\alpha$ kernel & magnetic field strength (G) & HXR association \\ \tableline
E1 &   260 & no  \\
E2 &   960 & yes \\
E3 &   240 & no  \\
E5 &   230 & no  \\
E6 &   220 & no  \\ \tableline 
W1 &  -500 & no  \\
W2 & -1050 & yes \\
W3 &  -260 & no  \\
W4 &  -300 & no  \\ \tableline

\end{tabular}
\end{center}
\end{table}

\clearpage

\end{document}